\documentclass[useAMS,usenatbib]{mn2e}
\usepackage{amssymb}
\usepackage{graphicx}
\usepackage{lscape}
\usepackage{longtable}
\usepackage{array}
\usepackage{natbib}
\bibpunct{(}{)}{;}{a}{}{,}

\newcommand\msun{$M_\odot$}

\newcommand\mMr{$m_\mathrm{max}$-$M_\mathrm{ecl}$ relation}
\newcommand\mecl{$M_\mathrm{ecl}$}

\newcommand\mmax{$m_\mathrm{max}$}
\title[The \mMr\,in NGC 4214]{Sampling methods for stellar masses and the \mMr\,in the starburst dwarf galaxy NGC 4214.}

\author[C.~Weidner et al.]
{Carsten Weidner$^{1,2}$\thanks{E-mail: cweidner@iac.es}, Pavel Kroupa$^{3}$\thanks{E-mail: pavel@astro.uni-bonn.de} and Jan Pflamm-Altenburg$^{3}$\thanks{E-mail: jpflamm@astro.uni-bonn.de}\\
$^{1}$Instituto de Astrof{\'i}sica de Canarias, Calle V{\'i}a L{\'a}ctea s/n, E38205, La Laguna, Tenerife,
Spain\\
$^{2}$Dept. Astrof{\'i}sica, Universidad de La Laguna (ULL), E-38206 La Laguna, Tenerife, Spain\\
$^{3}$Helmholtz-Institut f{\"u}r Strahlen- und Kernphysik (HISKP), Universit{\"a}t Bonn, Nussallee 14-16, D-53115 Bonn, Germany
}

\begin{document}
\bibliographystyle{aa}
\date{Received 2013 / Accepted 2014}

\pagerange{\pageref{firstpage}--\pageref{lastpage}} \pubyear{2014}

\maketitle

\label{firstpage}

\begin{abstract}
It has been claimed in the recent literature that a non-trivial relation between the mass of the most-massive star,~\mmax, in a star cluster and its embedded star cluster mass (the \mMr) is falsified by observations of the most-massive stars and the H$\alpha$ luminosity of young star clusters in the starburst dwarf galaxy NGC 4214. Here it is shown by comparing the NGC 4214 results with observations from the Milky Way that NGC 4214 agrees very well with the predictions of the \mMr\,and with the integrated galactic stellar initial mass function (IGIMF) theory. The difference in conclusions is based on a high degree of degeneracy between expectations from random sampling and those from \mMr, but are also due to interpreting \mmax\,as a truncation mass in a randomly sampled IMF.  Additional analysis of galaxies with lower SFRs than those currently presented in the literature will be required to break this degeneracy.
\end{abstract}


\section{Introduction}
\label{se:intro}

According to the integrated galactic stellar initial mass function (IGIMF) theory, the stellar initial mass function (IMF, see Appendix~\ref{app:IMF} for a description of the canonical IMF) of a whole galaxy needs to be computed by adding the IMFs of all newly formed star-forming regions. For galaxies with star formation rates (SFRs) smaller than 0.1 \msun/yr the IGIMF \citep[eq.~4.66 in][]{KWP13} is top-light with very major implications for the rate of gas consumption, when compared to the standard notion of an invariant IMF \citep{PK09}.

One fundamental corner stone of the IGIMF theory is the existence of a physical (aka non-trivial) relation between the mass of the most-massive star in a star cluster, \mmax, and the total stellar birth mass of the embedded star cluster, \mecl, which is called the \mMr. \citet{WK05b}, \citet{WKB09} and \citet{WKP13} quantified this relationship using resolved very young star clusters in the Milky Way and it was shown with high statistical significance that this relation leads to that the most-massive stars in star clusters are not as massive as would be expected if these clusters formed with their stars randomly drawn (for details on statistical sampling methods see \S~\ref{sub:sampling}) from the IMF. When assuming that the vast majority of star-formation occurs in causally connected events (embedded star clusters and associations) it is important to account for the initial distribution of these events, that is for the embedded cluster mass function (ECMF). Hence the IMF of a whole galaxy, the IGIMF, is the sum of all these events and the \mMr\,implies a suppression of the number of massive stars in galaxies with low star-formation rates (SFR). Because of the relation between the SFR and the mass of most-massive young star cluster, $M_\mathrm{ecl, max}$, in a galaxy \citep{WKL04}, galaxies with low SFRs tend to only form low-mass clusters and due to the \mMr\,only few massive stars. For average and large SFRs, however, the \mMr\,does not suppress the formation of massive stars and the integrated properties of the resulting stellar populations may then, at first sight, not be directly distinguishable from fully randomly sampled populations. 

The existence of the \mMr, however, is not without challenge and in a recent contribution \citet{ACC13} study a sample of unresolved young star clusters in HST images of the starburst dwarf galaxy NGC 4214. From the colours and H$\alpha$ fluxes and by deriving properties of these clusters via simulations the authors conclude that a physical most-massive-star--embedded-star-cluster relation, i.e. the \mMr, is ruled out and with it the theory of the IGIMF.

We also need to point out that the basic principle of the IGIMF is always true as the stellar population of any galaxy is the sum of all star-formation events in it. For more details on the IGIMF see \citet{WK05a} and \citet{KWP13} and more information on the \mMr\,can be found in \citet{WKP13}, \citet{GWKP12}, \citet{BKS11}, \citet{OK12} and \citet{KWP13}.

\begin{figure*}
\begin{center}
\includegraphics[width=8cm]{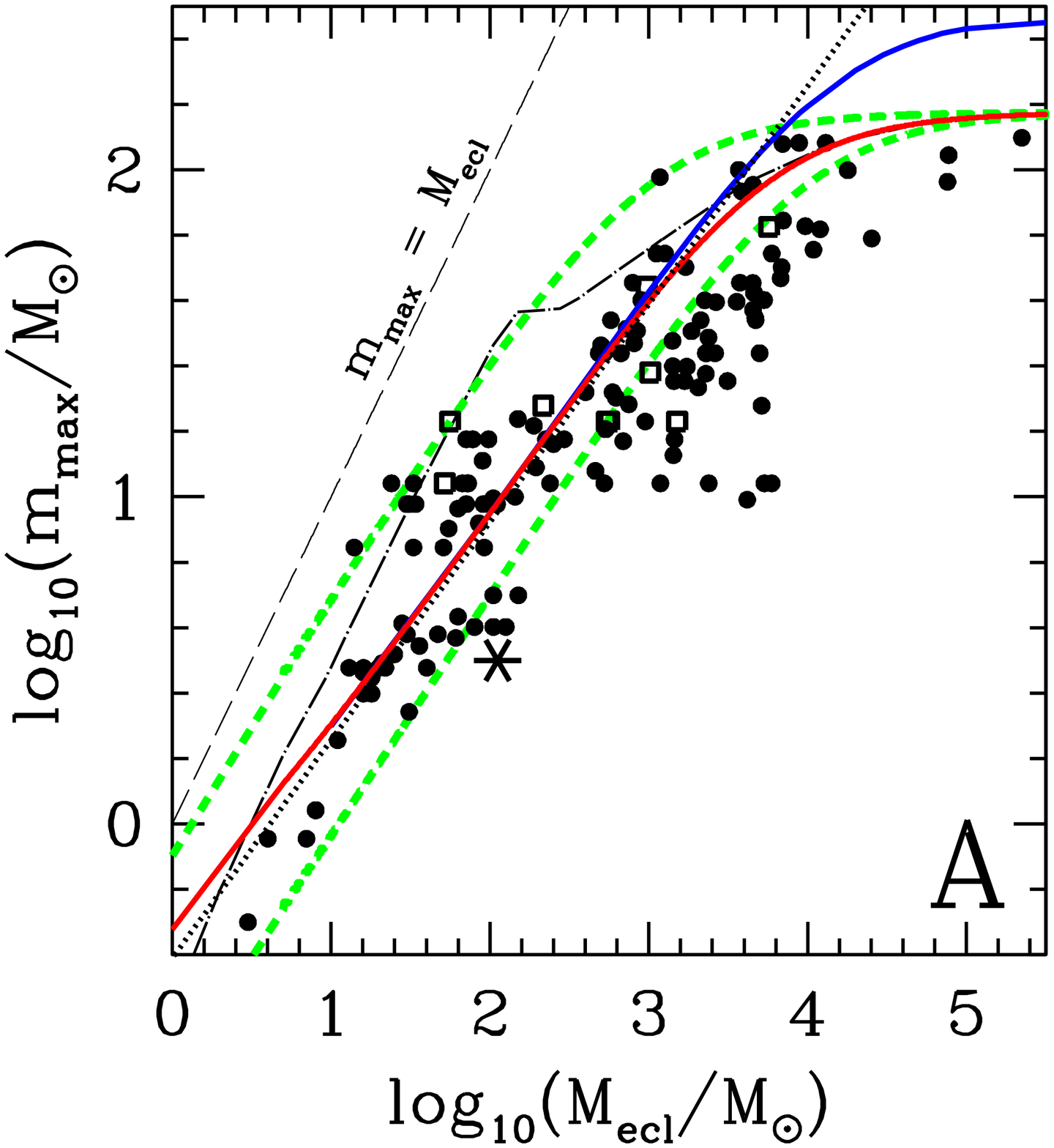}
\includegraphics[width=8cm]{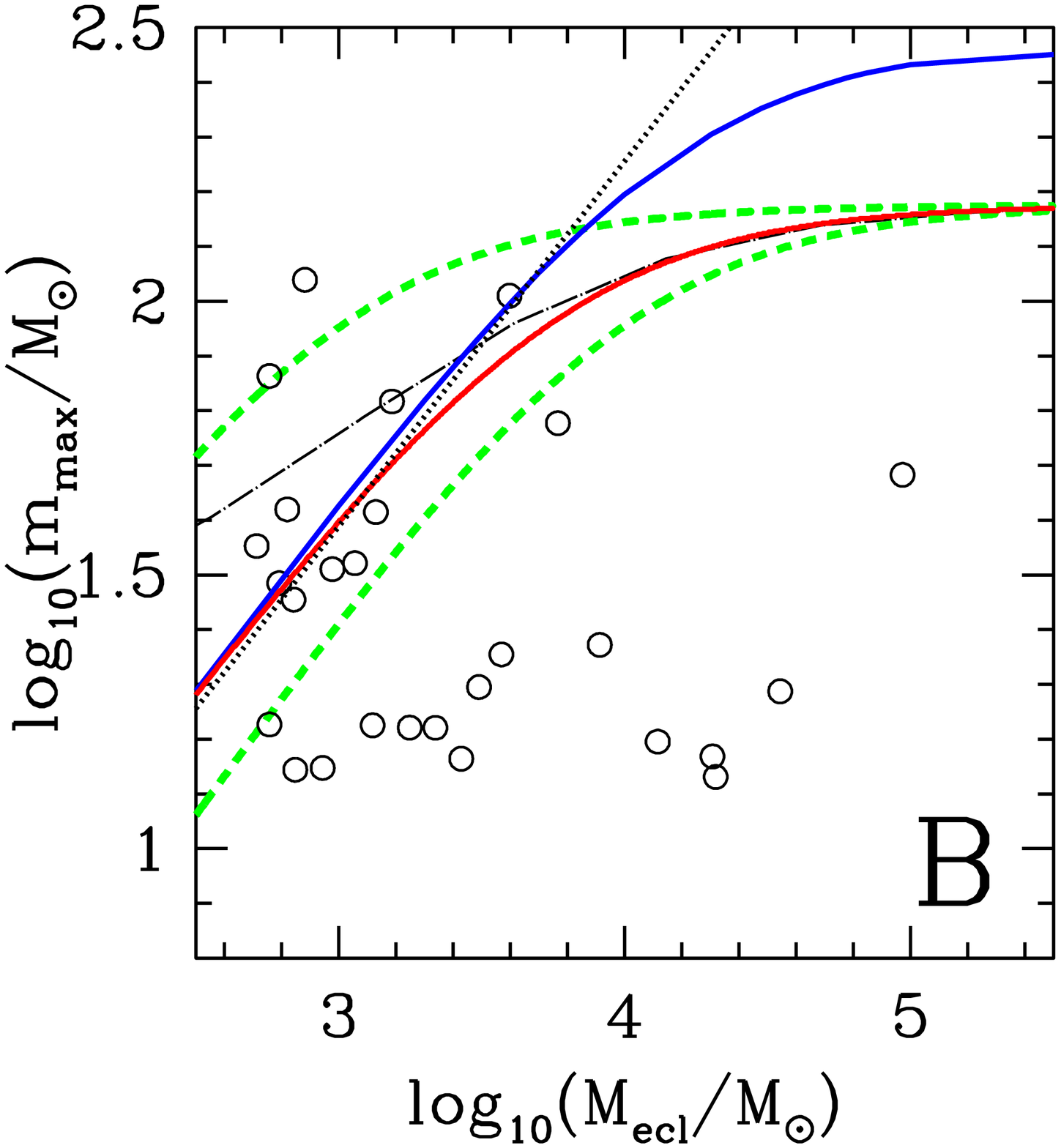}
\vspace*{-1.5cm}
\caption{Panel A: The mass of the most-massive star ($m_\mathrm{max}$) in an embedded cluster versus the stellar mass of the young dynamically un-evolved "embedded" cluster ($M_\mathrm{ecl}$). The filled dots are observations compiled by \citet{WKP13}. The boxes are mm-observations of massive pre-stellar star-forming regions in the Milky Way \citep{JSA09}. The solid lines through the data points are the medians expected for random sampling when using a fundamental upper mass limit, $m_\mathrm{max*}$, of 150\msun\,(lower grey solid line, red in the online colour version) and $m_\mathrm{max*}$ = 300\msun\,(upper grey solid line, blue in the online colour version). The dash-dotted line is the expectation value for random sampling derived from 10$^6$ Monte-Carlo realisations of star clusters. The change in slope at about \mecl\,= 100 \msun\,is caused by the fact that only below the fundamental upper mass limit ($m_\mathrm{max *}$ $\approx$ 150 \msun) it is possible to have clusters made of one star alone. Above this limit, also for random sampling clusters have to have several stars at least. This changes the behaviour of the mean \citep[for details see][]{SM08}. The dashed grey (green in the online colour version) lines are the 1/6th and 5/6th quantiles which would encompass 66\% of the \mmax\,data if they were randomly sampled from the canonical IMF (Fig.~\ref{fig:constrained}). The dotted black line shows the prediction for a relation by \citet{BBV03} from numerical models of relatively low-mass molecular clouds ($\le$ 10000 $M_\odot$). The thin long-dashed line marks the limit where a cluster is made out of one star. It is evident that random sampling of stars from the IMF is not compatible with the distribution of the data. There is a lack of data above the solid lines and the scatter of the data is too small despite the presence of significant observational uncertainties. The existence of a non-trivial, physical $m_\mathrm{max}$-$M_\mathrm{ecl}$ relation is implied. Panel B: Like panel A but shown as large open circles are the \mmax\,values from the modelling of NGC 4214 clusters by \citet{ACC13}. These values can not be directly compared with the direct measurements shown in panel A as they are the results of best-fits of unresolved cluster photometry with models.} \label{fig:mmaxmecl}
\end{center}
\end{figure*}

The \mMr\,has been derived from theoretical arguments as well as observational data in the Milky Way and the Magellanic Clouds. As can be seen in panel A of Fig.~\ref{fig:mmaxmecl}, it shows that star clusters are depressed in the formation of massive stars significantly more so than is expected from random sampling from a stellar initial mass function (IMF). {\it We emphasise that all available data on very young populations have been used and the selection criteria are only one of age being younger than 4 Myr and no supernova remnants must be in the cluster.}

In panel B of Fig.~\ref{fig:mmaxmecl} the \mmax\,values of \citet{ACC13}, which have not been published but were kindly provided by Daniela Calzetti (priv. communication), are plotted. These \mmax\,values are not based on direct measurements but are the results of best-fits of photometric data of the clusters in NGC 4214 with cluster models and are therefore not plotted together with the data of panel A. While the resolved cluster data into individual stars show a strong trend of rising \mmax\,with increasing \mecl\,the \citet{ACC13} data form a flat distribution. This is very surprising as even in the case of fully randomly sampling stars from the IMF a trend with \mecl\,is expected. Therefore, the NGC 4214 data have a trend which is hidden in the (unknown) error bars or the clusters in NGC 4214 are incompatible with any known sampling procedure.

The cluster mass axis in panel B of Fig.~\ref{fig:mmaxmecl} has been limited to \mecl\,$\ge$ 300\msun\,as only such objects are subject of the \citet{ACC13} paper. This also removes the need for a deeper discussion of massive stars allegedly formed in isolation as no stars with stellar mass above 300\msun\,are known and therefore these are not needed to be taken into account in the debate about random sampling in star clusters. For a detailed discussion of recent claims of O stars formed in isolation see \citet{GWKP12}.

The actual physical existence of a \mMr\,is not subject of this publication. Instead, we critically discuss the way the \mMr\,has been applied in \citet{ACC13} as well as by \citet{FDK11}, \citet{DFK12} and others. We show these applications to be problematic because they are not self-consistent and because they do not reproduce the input \mMr. In \S~\ref{se:why} it is shown why the \mMr\,can not be a truncation limit. And in \S~\ref{se:ngc4214} the NGC 4214 cluster data are compared with the \citet{WKP13} sample of star clusters before the results are discussed in \S~\ref{se:conclusions}.

\section{Why the \mmax-\mecl\,relation is not a truncation limit}
\label{se:why}

\subsection{Sampling methods}
\label{sub:sampling}

\begin{figure*}
\begin{center}
\includegraphics[width=16cm]{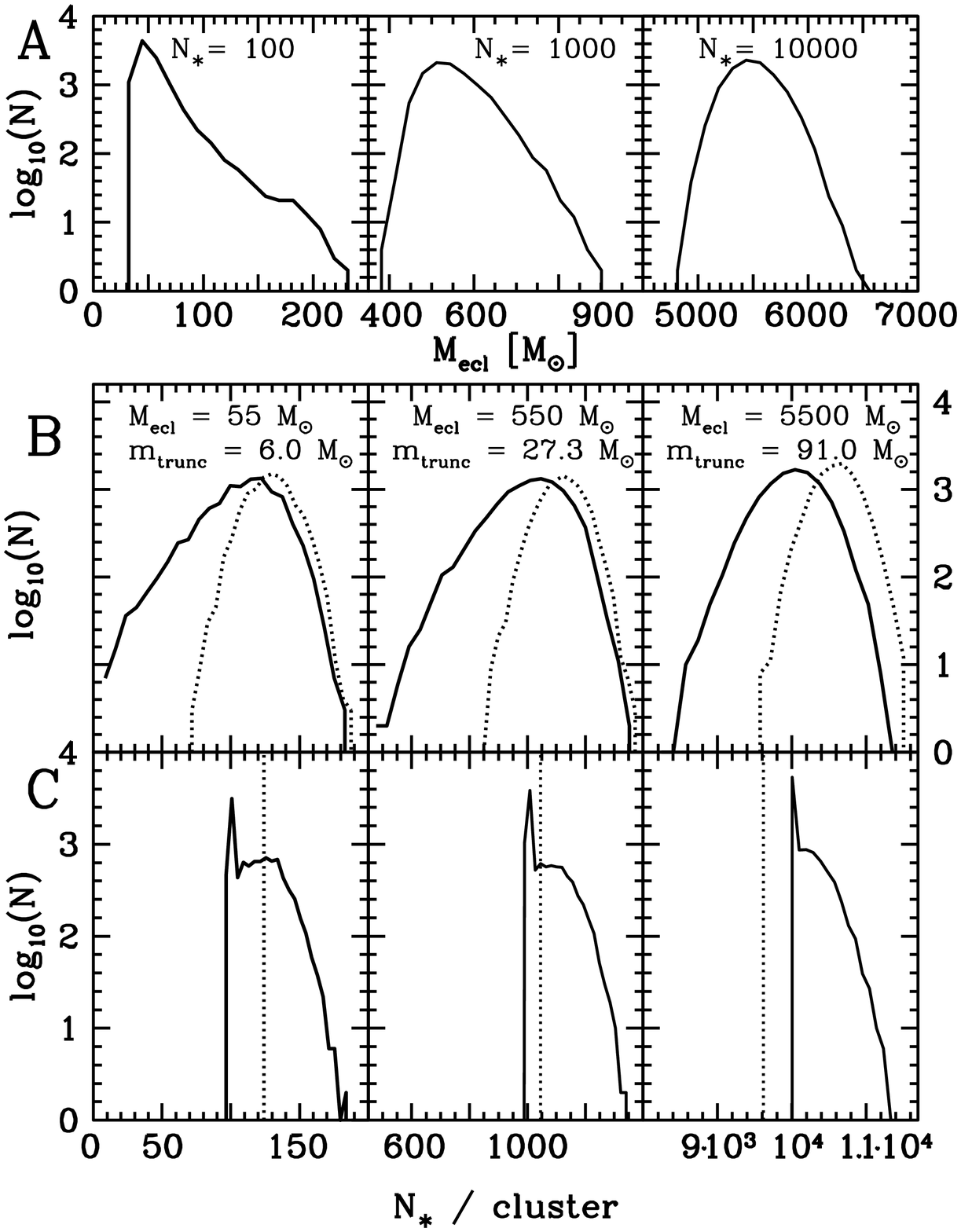}
\caption{The impact of the different sampling methods as described in \S~\ref{sub:sampling} on the cluster properties. In the uppermost row (A) three different $N_\ast$ are randomly sampled 10000 times from the IMF and the resulting \mecl\,are plotted against how often they are realised. Beginning from the left, $N_\ast$ = 100, $N_\ast$ = 1000 in the middle and $N_\ast$ = 10000 on the right. In row B and C, three \mecl\,(from left to right, \mecl\,= 55 \msun, \mecl\,= 550 \msun\,and \mecl\,= 5500 \msun) are sampled 10000 times using constrained sampling without truncation (solid lines in row B), constrained sampling with a truncation (dotted lines in row B), sorted sampling (solid lines in row C) and optimal sampling (dotted lines in row C). Plotted are how often the resulting number of stars per cluster are realised, binned in 25 bins between the least and most-massive cluster (row A) and 25 bins between the cluster with lowest and highest $N_\ast$ (B \& C).} \label{fig:sampling}
\end{center}
\end{figure*}

\begin{figure*}
\begin{center}
\includegraphics[width=16cm]{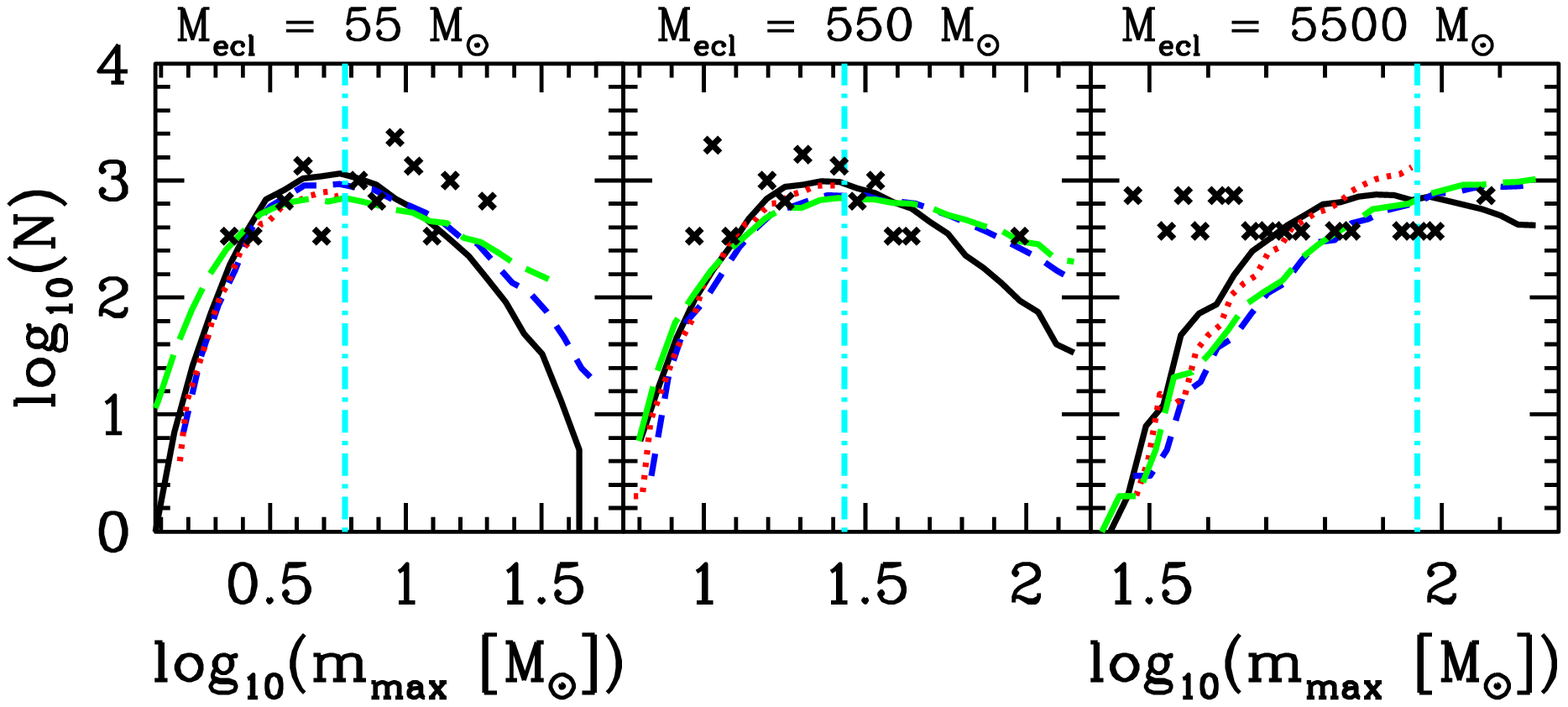}
\vspace*{-14cm}
\caption{The distributions of the mass of the moss-massive star, \mmax\,for different sampling methods for three different cluster masses of (from left to right) \mecl\,= 55 \msun, \mecl\,= 550 \msun\,and \mecl\,= 5500 \msun. For the solid line, sorted sampling was used, while the short-dashed (blue) line uses constrained sampling, long-dashed (green) line uses random sampling and the dotted (red) line constrained sampling with the \mMr\,as a truncation limit. The vertical dash-dotted line (cyan) is the \mmax\,value for optimal sampling for each given cluster mass. The crosses indicate observed clusters from the \citet{WKP13}-sample for which the target masses are within the errors bars and scaled to 10000 clusters. For the left and the middle panel 30 clusters each have the corresponding mass within their error bars and for the right panel 27 clusters.} \label{fig:samplingmmax}
\end{center}
\end{figure*}

Before it is shown in \S~\ref{sub:mmr} that using the \mMr\,as a truncation limit for populating star clusters with stars with a Monte-Carlo method is wrong when trying to preserve the \mMr\,in the process, a range of different sampling methods of stars from an IMF and their important differences are described as these differences have a strong impact on numerically created stellar populations:

\begin{itemize}
\item Random sampling\\
In order to apply real random sampling for creating numerical star clusters a number of stars, $N_\ast$, has to be chosen. This number can be fixed, random or itself taken from a distribution. This number of stars, $N_\ast$, is then randomly taken from the IMF in order to arrive at a distribution of stellar masses. Adding up these stellar masses results in \mecl. Row A of Fig.~\ref{fig:sampling} shows the resulting distribution of cluster masses for 10000 Monte-Carlo realisations of three different $N_\ast$. On the left is the case $N_\ast$ = 100 stars, in the middle $N_\ast$ = 1000 stars and on the right $N_\ast$ = 10000 stars. The figure shows that the resulting cluster mass for a given $N_\ast$ can vary strongly. Especially for relatively low $N_\ast$, \mecl\,can differ by a factor of more than three. The distribution of \mmax\,for the 10000 Monte-Carlo realisations are shown as long-dashed (green) lines for the different sample sizes in Fig.~\ref{fig:samplingmmax}.\\

\item (Mass-) Constrained sampling\\
Often, not a number of stars is initially available for a given problem but the cluster mass \mecl\,is the physically relevant quantity. This should then be referred to as (mass-)constrained sampling. To reach \mecl\,stars are randomly chosen from the IMF and their masses are added. When to finish this process is handled differently by different authors and results in severe discrepancies. Usually the process is stopped when the sum of the chosen stars is larger than \mecl\,but one can then keep the last star or remove it from the sum and therefore either consistently over- or under sample the target mass. Sometimes also a criterion is used to decide whether or not the last star should be kept in the sample. For example, the last star drawn is kept if the \mecl\,is only exceeded by a pre-set accuracy, e.g.~$\epsilon$ $\approx$ 10\%, or what mass is closer to \mecl\,percentage wise. Alternatively, if the last star does not satisfy a given accuracy it is replaced with stars drawn from the IMF until \mecl\,fits the accuracy. A sub-method of constrained sampling would be to use the \mMr\,to set the \mmax\,for a given \mecl\,and discard any star more massive than this \mmax. Note that either way how the final star is handled, constrained sampling always changes the input IMF as stars randomly drawn from the IMF are discarded in order for the sum of the masses of the stars to represent the chosen input \mecl. In Fig.~\ref{fig:sampling} in row B three examples of constrained sampling are shown by plotting the number of stars per cluster for 10000 Monte-Carlo realisations. On the left side of the row the Monte-Carlo results are shown for \mecl\,= 55 \msun, in the middle for \mecl\,= 550 \msun\,and on the right for \mecl\,= 5500 \msun. These masses are chosen so that on the average the number of stars per cluster is 100, 1000 and 10000, respectively, to arrive at similar clusters as in row A of the Figure. The solid lines in row B of Fig.~\ref{fig:sampling} refer to constrained sampling without any limits while the dotted lines use the \mMr\,as a truncation limit for the most-massive star for a given cluster mass (ie., a star is discarded if its mass lies above \mmax\,for the respective pre-defined \mecl\,value). Because the \mMr\,deviates stronger from the expectations of random sampling for larger \mecl, the dotted and solid lines diverge more for larger \mecl\,as well. In Fig.~\ref{fig:samplingmmax} the distribution of the \mmax\,values from constrained sampling are plotted with short-dashed (blue) lines, while for constrained sampling with the \mMr\,as the truncation limit, dotted (red) lines are used. Note that introducing a truncation limit for constrained sampling changes the distribution of number of stars per cluster and the distribution of the \mmax\,values significantly {\it especially} for more massive clusters. As for constrained sampling the aim is to fit the target \mecl\,as well as possible, stars sampled from the IMF are discarded and therefore the IMF is changed in this process and using a truncation limit amplifies this effect. {\it Hence, constrained sampling should not be confused with random sampling}.\\

\item Sorted sampling \citep{WK05b}\\
Sorted sampling is more complex. Here, the given \mecl\,is divided by the mean mass, $\overline{m}$, of the input IMF\footnote{The canonical IMF, for example, has $\overline{m}$ $\approx$ 0.55\msun\,between 0.08 and 150\msun.}. This results in an expected number of stars, $N_\mathrm{expect}$, for that cluster with the input IMF. This $N_\mathrm{expect}$ is then randomly taken from the IMF and sorted by mass. Starting from the lowest mass star the stellar masses are added and compared with \mecl. If the sum is larger than \mecl\,massive stars are removed until the sum is within 10\% of \mecl. However, if the sum is smaller than \mecl, the difference between \mecl\,and the sum is calculated and this difference then divided by $\overline{m}$. This results in an additional number of stars which are randomly taken from the IMF. The initial $N_\mathrm{expect}$ number of stellar masses and this additional number of stellar masses is then together sorted by mass and summed up. This procedure is repeated until \mecl\,is reached to within 10\% accuracy. Row C of Fig.~\ref{fig:sampling} shows the distribution of number of stars per cluster for the same masses as in row B. Due to the properties of sorted sampling the number of stars per cluster is sharply peaked at the nominal number given by the cluster mass and the $\overline{m}$ of the chosen IMF. The disadvantage of sorted sampling is that the distribution of the number of stars per clusters (row C in Fig.~\ref{fig:sampling}) has a very sharp edge at the lower end close to the expected number of stars per cluster at a given \mecl. However, the distribution of the \mmax\,values for this sampling method (solid lines in Fig.~\ref{fig:samplingmmax}) is broader and more similar to other sampling methods.\\

\item Optimal sampling \citep{KWP13}\\
Optimal sampling assumes that star formation is deterministic, e.~g., that with the exact same initial conditions the resulting clusters would be identical. For optimal sampling the stellar masses in a cluster with a given \mecl, \mMr\,and IMF are analytically determined by taking \mmax\,from the \mMr\,and then calculating iteratively the next star using the IMF. This sampling method therefore always reaches \mecl\,exactly and always with the same number of stars with the same masses for a given \mecl\,and statistical variations of the final discretised IMF are eliminated altogether \citep[for the analytical and numerical implementation of optimal sampling see][]{KMK11}. The number of stars per cluster when using optimal sampling is shown in Fig.~\ref{fig:sampling} as a dotted line in row C and is constant for a given \mecl. Optimal sampling constitutes an extreme interpretation of the star formation process as being completely deterministic through perfect self-regulation. In Fig.~\ref{fig:samplingmmax} the \mmax\,values for the three cluster masses are shown as vertical dash-dotted (cyan) lines. While the \mmax\,value for optimal sampling is identical to the truncation limit for constrained sampling with the \mMr\,as a limit (dotted lines), optimal sampling never under-samples (or over-samples) the \mMr.
\end{itemize}

The differences in the resulting stellar populations of star clusters due to the different sampling methods as shown in Figs.~\ref{fig:sampling} and \ref{fig:samplingmmax} should clearly indicate that populating star clusters with a Monte-Carlo method is not as straightforward and trivial as it might seem and a clear terminology and description of the used procedures is vital. An important issue of constrained sampling with the \mMr\,as a truncation limit, as seen in Fig.~\ref{fig:samplingmmax}, is that this sampling method never can produce stars with a mass above the truncation. Other sampling methods are either always on the limit (optimal sampling) or all \mmax\,values distribute around the limit and henceforth do not bias the resulting \mmax\,values to be preferably below the truncation. Observed clusters in the Milky Way (crosses) as well distribute above and below truncation limit, hinting again that using the \mMr\,as a truncation is unphysical. Furthermore, it is not clear what might be a 'preferred' sampling method for actual stars formed in molecular clouds as the existence of a non-trivial \mMr\,casts doubt on a random sampling process. A result obtained in \citet{WKP13} was that the measurement uncertainties in mmx and in \mecl\,values appear to account for most of the scatter in the \mMr\,diagramm such that the hypothesis that there is no true physical dispersion, i.e.\,that optimal sampling may be correct, cannot be excluded.

\begin{figure*}
\begin{center}
\includegraphics[width=8cm]{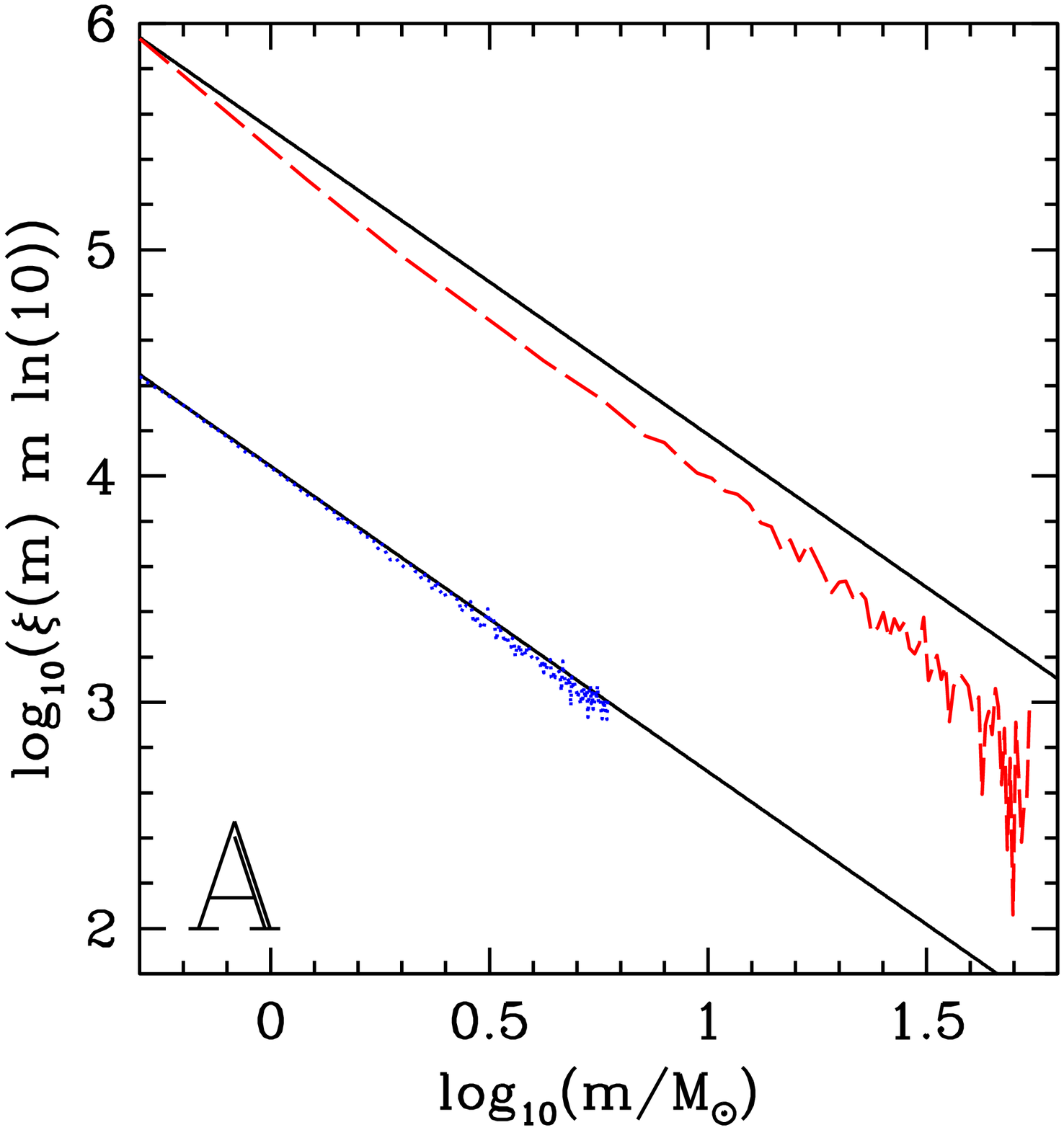}
\includegraphics[width=8cm]{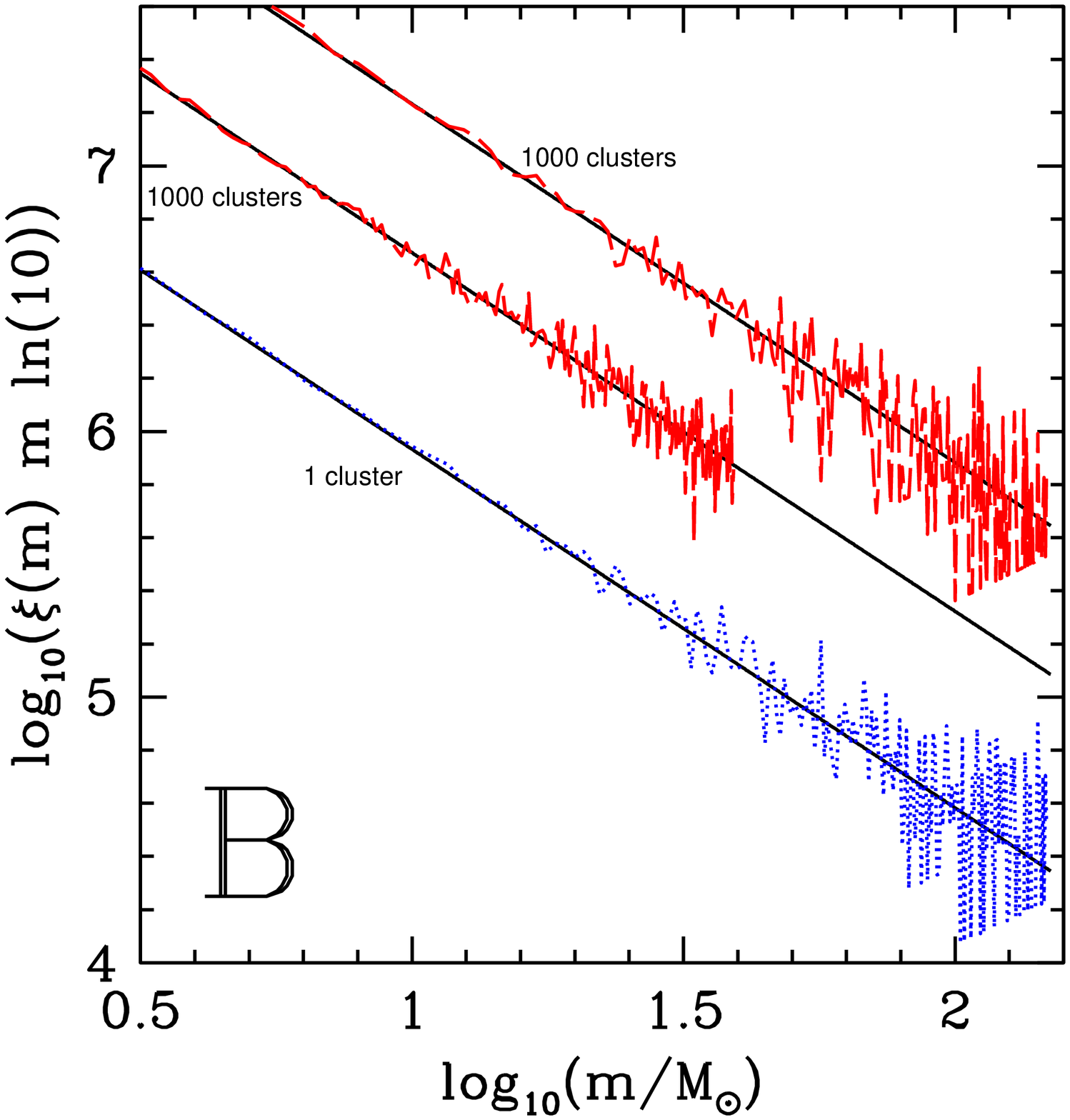}
\vspace*{-1.5cm}
\caption{Panel A: The impact of constrained sampling on the IMF of low-mass clusters. The (red) dashed line is the IMF of 10000 clusters with \mecl\,= 55\msun\,populated from the canonical IMF by constrained sampling, while for the dotted (blue) line the 10000 clusters were populated with constrained sampling but using the \mMr\,as a truncation limit. The two solid (black) lines indicate the canonical Salpeter slope of $\alpha_3$ = 2.35. The IMFs are arbitrarily shifted vertically to enhance clarity. Note the different slopes. Panel B: The IMFs of three different scenarios. The lower (blue) dotted line, labelled '1 cluster', is the IMF of a star cluster with \mecl\,= 10$^6$ \msun\,made by constrained sampling, while the two upper (red) dashed lines are the arbitrarily shifted IMFs of  two times 1000 clusters each with \mecl\,= 1000 \msun. In the case of the top-most line constrained sampling was used while in the case of the lower dashed line again constrained sampling was used but with the analytical \mMr\,as a truncation limit. The three solid (black) lines indicate the Salpeter slope of $\alpha_3$ = 2.35. No apparent difference in the three IMFs is visible other than that in the case of the use of the \mMr\,as a truncation limit, the IMF is not sampled for m $>$ \mmax\,= 39.1 \msun.} \label{fig:IMFs}
\end{center}
\end{figure*}

Besides the difference in typical cluster mass or number of stars per cluster, the choice of the sampling method has other implications for the numerically made stellar populations. Panel A of Fig.~\ref{fig:IMFs} shows the summed IMFs of 10000 Monte-Carlo clusters with \mecl\,= 55 \msun\,sampled from the canonical IMF. For the (red) dashed line, constrained sampling as described above was used while for the (blue) dotted line also constrained sampling was chosen but the stellar masses in the clusters are limited to lie below the \mmax\,value for \mecl\,= 55 \msun. The two solid lines are vertically arbitrarily shifted IMFs with a slope of $\alpha_3$ = 2.35. While the truncated sampled clusters reproduce the canonical IMF very well, the clusters using constrained sampling (usually called 'random' sampling) change the IMF quite significantly. This steepening of the IMF only impacts clusters below or close to the fundamental upper mass limit for stars (here set to 150 \msun) because in order to reasonably reach the targeted cluster mass often massive stars have to be removed from the cluster.

Note that no sampling method has been found yet which reproduces the observed \mMr\,particularly well. But optimal sampling and sorted sampling are exact fits for the analytical \mMr. Additionally, many more variations of the above mentioned sampling methods are possible and in use and completely different ones are also possible. This has to be kept in mind as comparing the results of different sampling methods is not straightforward and can be misleading.

\subsection{The \mmax-\mecl\,relation}
\label{sub:mmr}

\begin{figure}
\begin{center}
\includegraphics[width=8cm]{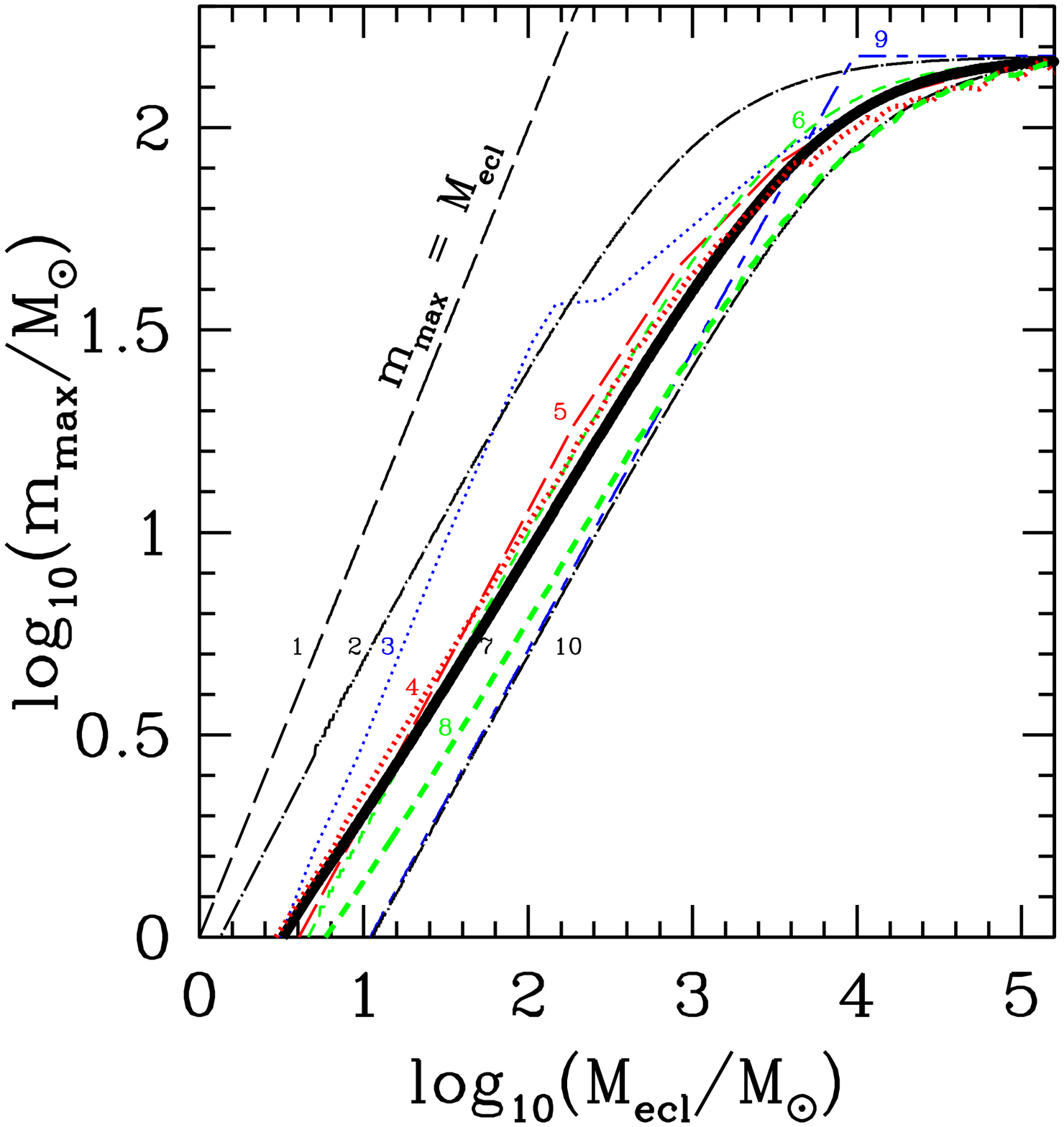}
\vspace*{-1.5cm}
\caption{A selection of different \mMr s. The lines denote the following relations (as described in \S~\ref{sub:sampling}) by number. 1 (long-dashed line): the limit where a cluster is made out of one star only. 2 (dash-dotted line): the upper 66.6\% line. Between this line and line number 10, 2/3rds of all the \mmax\,values  should lie if the stars are randomly sampled from the canonical IMF with an upper limit of $m_\mathrm{max \ast}$ = 150 \msun. 3 (blue dotted line): the mean, trivial \mMr\, for random sampling. 4 (red dotted line): the \mMr\,determined from 10$^6$ Monte Carlo clusters using sorted sampling. 5 (red long-dashed line): the \mMr\,for constrained sampling. 6 (green short-dashed line): the median \mMr\,for random sampling. 7 (thick black solid line): the analytical \mMr\,(eqs.~\ref{eq:mmr1} and \ref{eq:mmr2}). 8 (green short-dashed line): the \mMr\,arrived at when using constrained sampling and the analytical \mMr\,as a truncation limit. 9 (blue long-dashed line): the mode \mMr\,for random sampling. 10 (dash-dotted line): the lower 66.6\% line. Between this line and line number 2, 2/3rds of all the \mmax\,in clusters should lie if the stars are randomly sampled from the canonical IMF with an upper limit of $m_\mathrm{max \ast}$ = 150 \msun.} \label{fig:trunc}
\end{center}
\end{figure}

Before discussing the issue of truncation and the use of the \mMr\,we first need to define {\it which} \mMr\,is actually meant. When using random sampling, as discussed in \S~\ref{sub:sampling}, but using the masses of the resulting star clusters and comparing them to their most massive stars, a \mMr\,is also observed. It is generally called the trivial \mMr. Constrained sampling results in a slightly different trivial \mMr\,while the physically interesting ones are the analytical \mMr, as deduced in \citet{WK04}, and the observed (empirical) \mMr, which was quantified more precisely in \citet{WKP13}. Additionally, for the relations derived from Monte-Carlo sampling methods, the distinction has to be made between the mean \mMr, which uses the mean value of millions of most-massive stars in cluster mass bins, the median \mMr\, which uses the median of the \mmax\,values and the mode \mMr, for which the mode (peak = most common) values of \mmax\,is used. As the IMF is a non-symmetric function, all three \mMr s are different. Also useful are the upper and lower relation between which 66.6\% of all the clusters are expected to lie. Because in \citet{ACC13} the analytical \mMr\,is used, we focus on this one as well. Some of the different \mMr s are shown in Fig.~\ref{fig:trunc}. Of course, all \mMr s constructed from the IMF change when using a different IMF and/or different lower and upper mass limits for this IMF.

The analytical \mMr\,can be derived by numerically solving the following system of two equations. The first one describes  how the mass of a cluster, \mecl, is derived from the IMF, $\xi(m)$,
\begin{equation}
\label{eq:mmr1}
M_{\rm ecl} = \int_{m_{\rm low}}^{m_{\rm max}}m \cdot \xi(m)~dm,
\end{equation}
The canonical IMF is described in appendix~\ref{app:IMF} and $m_\mathrm{low}$ and $m_\mathrm{max}$ are, respectively, the lower and the upper mass limit of the IMF. We employ $m_\mathrm{low}$ = 0.08 \msun\,while \mmax\,is the value intended to be calculated for a given \mecl.\\

The second equation states that there is exactly one most-massive star in a cluster,
\begin{equation}
\label{eq:mmr2}
1 = \int_{m_{\rm max}}^{m_{\rm max *}}\xi(m)~dm,
\end{equation}
with $m_\mathrm{max *}$ being the fundamental upper mass limit for stars.

The analytically derived \mMr\, has been unfortunately interpreted as a truncation limit or a randomly sampled IMF by \citet{FDK11}, \citet{DFK12} and others. It can be seen in Fig.~\ref{fig:samplingmmax} and Fig.~\ref{fig:trunc} that this assumption of the \mMr\,being a truncation limit leads to significant differences in the resulting artificial populations.

The black solid line in Fig.~\ref{fig:trunc} shows the analytical \mMr\,(eqs.~\ref{eq:mmr1} and \ref{eq:mmr2} and equation 10 in \citealt{PWK07}), while the (red) dotted line (4) is the \mMr\,derived from sorted sampling \citep{WK05b}. To derive this line, 10$^6$ star clusters were constructed by sorted sampling and their most-massive star and cluster mass was recorded. These data were then used to calculate the mean most-massive star within a range (ie. bin) of cluster masses. Clusters with masses within 10\% of the aimed-for cluster mass have been used. The short-dashed line is arrived at when using the analytical \mMr\,as a truncation limit when constructing 10$^6$ clusters with mass-constrained sampling \citep{WK05b} and calculating the \mMr\,for this experiment in the same way as for the clusters generated using sorted sampling. As the mean most-massive star derived from Monte-Carlo experiments with the analytical \mMr\,as a truncation limit does not reproduce the analytical \mMr\,(the mean value always lies below the analytical \mMr), it is obvious that this truncation can not be the right procedure to implement the \mMr\,into Monte-Carlo star cluster populations. Using the analytical or median \mMr\,as a truncation limit such that only stars with masses below this relation are allowed in random sampling to be present thus leads to a significant underestimate of the masses of the most massive stars in the population.

In order to quantify this effect, the distances of the 10$^6$ Monte-Carlo star clusters, of the observed sample of clusters of \citet{WKP13} and of the clusters in NGC 4214 to the analytical \mMr\,are calculated using\\
\begin{eqnarray}
\label{eq:dist}
\lefteqn{ {\rm distance}_i = \min(\sqrt{[\log_{10}(m_\mathrm{max, i}) - \log_{10}(m^{'}_\mathrm{max})]^2 +}}
\nonumber \\
&&{}\overline{+ [\log_{10}(M_\mathrm{ecl, i}) - \log_{10}(M^{'}_\mathrm{ecl})]^2}),
\end{eqnarray}
where $m_\mathrm{max, i}$ and $M_\mathrm{ecl, i}$ are, respectively, the \mmax\,and the $M_\mathrm{ecl}$ values of the i-th data point and $m^{'}_\mathrm{max}$ and $M^{'}_\mathrm{ecl}$ are, respectively, the \mmax\,and the $M_\mathrm{ecl}$ values of the analytical \mMr. The distances of a given $m_\mathrm{max, i}$ and $M_\mathrm{ecl, i}$ couple are calculated to all points of the analytical \mMr, and the smallest value is taken as the final distance. When the \mmax\,value is lower than the \mmax\,value of the analytical \mMr\,for the given \mecl,the distance is multiplied by -1.

When using the \mMr\,as a truncation limit for constrained sampling, no clusters with most-massive stars above the \mMr\,are possible, while sorted sampling as well as the observations in the MW and LMC can be clearly found above (and below) the analytical \mMr. Thus, making star clusters with such a truncation does not reproduce the observed variety of massive stars in star clusters. {\it By design, this truncation can not reproduce the input \mMr\,which itself is based on observational constrains}. In order to be able to do so, the resulting \mmax\,values need to have a similar spread around the analytical \mMr\,as the observations have.

Additionally, it can be seen in Fig.~\ref{fig:mmaxmecl} that only 11 of the 27 (41\%) most-massive stars in the clusters of \citet{ACC13} are between the two dashed lines, while for random sampling 2/3rd of the most-massive stars should be in this region. The NGC 4214 clusters are {\it not} compatible with random sampling.

\subsection{The connection between the \mMr\,and the IGIMF}
\label{sub:connection}

\begin{figure}
\begin{center}
\includegraphics[width=8cm]{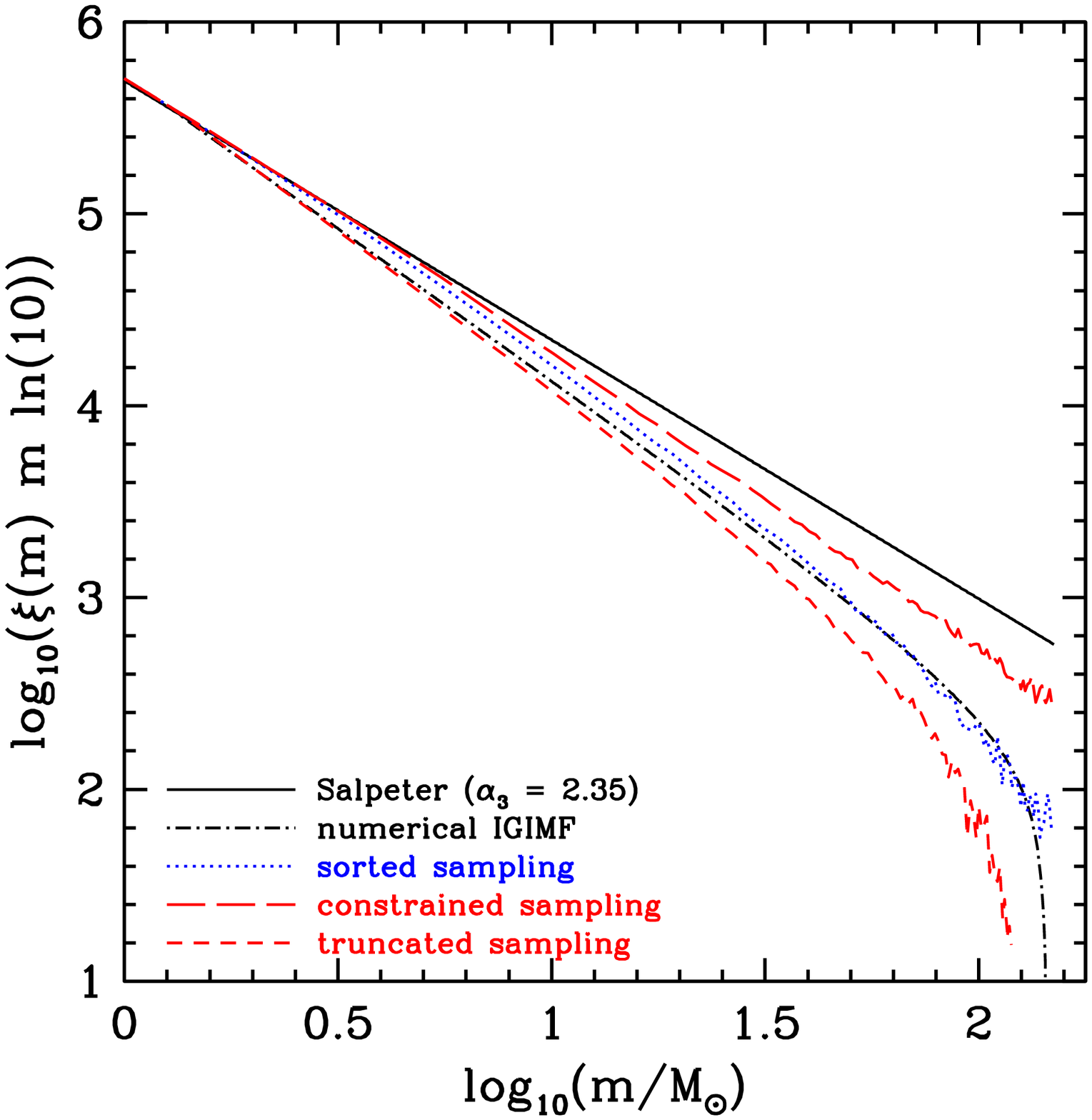}
\vspace*{-1.5cm}
\caption{IGIMFs \citep[see eq.\,4.66 in][]{KWP13} resulting when using different sampling methods of stars in star clusters \citep{WK05b}. For the solid line random sampling was used (identical to the canonical IMF), for the red long-dashed line mass-constrained sampling was used, for the blue dotted line sorted sampling was employed, while the dash-dotted line shows the numerical solution for IGIMF and for the red short-dashed lines the \mMr\,was used as an truncation limit. Note how the different sampling methods change the resulting IGIMF considerably. Only sorted sampling reproduces the analytically derived IGIMF. Below 1 \msun\,the IMFs plotted here agree with the canonical IMF (Appendix~\ref{app:IMF}). In all cases the embedded cluster mass function was assumed to be a power-law with a slope of 2.0 between 5\msun\,and 10$^6$\msun.} \label{fig:constrained}
\end{center}
\end{figure}

While the \mMr\,is an important constraint on star-formation models, it also has far reaching consequences for the stellar populations of galaxies. Only the \mMr\,resulting from (pure) random sampling keeps the IMF scale-free. This means that any superposition of IMFs from different star-forming regions will result in the same IMF for the whole galaxy (the IGIMF) as it is observed in individual star clusters. Any other of the here discussed \mMr s, or equivalently sampling methods, results in breaking this scale-free behaviour. Even for (mass-) constrained sampling (which is often labeled random sampling), 1000 star-forming regions of 100\msun\,do not result in the same IMF as one region with 10$^5$\msun. For constrained sampling the reason is obvious because a star-forming region of 100\msun\,will never be able to form a star above 100\msun.
The different IGIMFs for several sampling methods are visualised in Figure~\ref{fig:constrained}. For the solid line pure random sampling was used and it results in the same IMF as the input IMF. The other lines use different sampling methods to populate clusters with stars. For the long-dashed line constrained sampling was applied, for the dotted line sorted sampling and for short-dashed line the analytical \mMr\,was used as a truncation limit while the clusters were filled with stars using constrained sampling. The clusters were Monte-Carlo sampled under the assumption that 100\% of all stars form in embedded clusters, i.e.~no isolated star-formation occurs.

\section{The NGC 4214 data}
\label{se:ngc4214}

NGC 4214 is an irregular dwarf starburst galaxy at about 3 Mpc distance with a SFR of 0.16\msun\,yr$^{-1}$ deduced from the H$\alpha$ flux and 0.22\msun\,yr$^{-1}$ from its UV flux \citep{ACC13}. For such a high SFR, the IGIMF theory indeed does not predict any strong reduction in the number of OB stars compared to the canonical IMF and therefore no significant differences to random sampling are expected. This is because, using the SFR-$M_\mathrm{ecl max}$-relation \citep{WKL04} for a SFR of 0.2 \msun\,yr$^{-1}$ and the analytical \mMr\,\citep[eq.~8 in][]{WK04}, an upper stellar mass limit for the whole galaxy of 130\msun\,is to be expected, which is very well consistent with the observations by \citet{ACC13}. This can also be seen from figure~5 in \citet{PWK07} where a difference between SFRs determined by UV and H$\alpha$ fluxes is only significant for SFRs below 0.05 \msun\,yr$^{-1}$. It should be noted here that in \citet{ACC13} the authors themselves write "Specifically galaxies with star formation rates (SFR) below the threshold for which IMF variances have been suggested ($\le 0.1 M_\odot~\mathrm{yr}^{-1}$) need to be investigated". Therefore, NGC 4214 is ill-suited to study the IGIMF and the \mMr. Generally speaking, integrated properties are unsuitable tools to study subtle differences in the IMF of stellar populations. \citet[][table~1 and 2]{WKP13} show that the expected numbers of A, B and O stars from clusters with masses as low a 10 \msun\,are indistinguishable when using random and sorted sampling. Though, only the latter reproduces the analytical \mMr.

\begin{figure}
\begin{center}
\includegraphics[width=8cm]{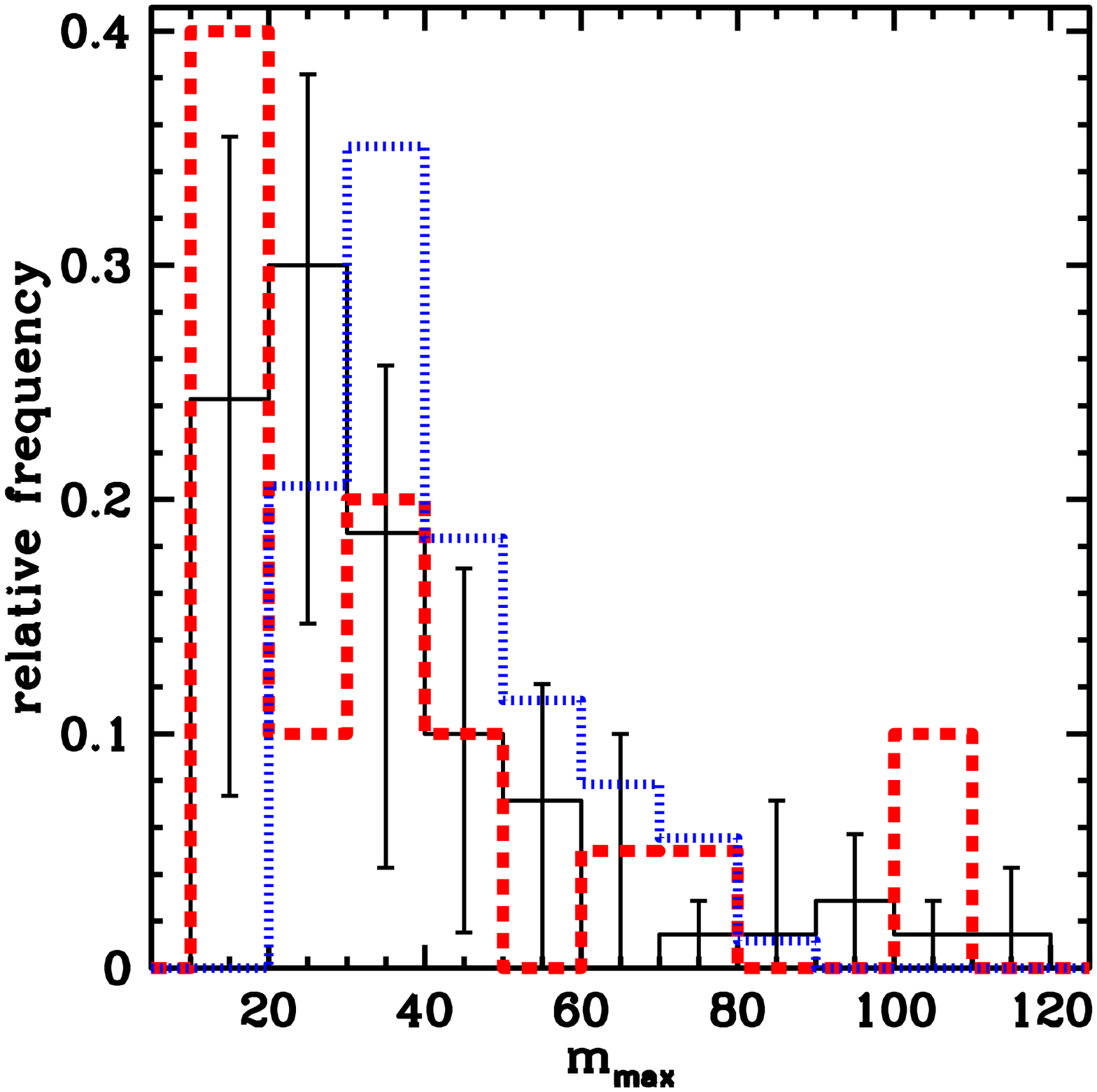}
\vspace*{-2.0cm}
\caption{The dashed (red) histogram shows the 'observationally' derived relative frequency of the most-massive star mass, $m_\mathrm{max}$, for the NGC 4214 clusters with masses around $10^3$ $M_\odot$ from \citet{ACC13}. The solid (black) histogram is the same for Milky Way star clusters in the mass range 500 to 4000\msun\,from the cluster data from table A1 in \citet{WKP13}. The error bars for the Milky Way data are calculated from the errors in $m_\mathrm{max}$ and the Poisson error for the number of clusters per bin. The dotted (blue) histogram is the result of a Monte-Carlo simulation which uses the analytical \mMr\,\citep{PWK07} to assign \mmax\,to 10$^6$ star clusters randomly drawn from a cluster mass function.} \label{fig:mmaxfreq}
\end{center}
\end{figure}

\citet{ACC13} apply the SLUG code \citep{DFK12} to derive masses and ages of the clusters and their most-massive stars in NGC 4214. It must be noted here that these masses determined for \mmax\,are purely model results as the observations do not resolve individual stars. Whether such model results are bijective (have only one singular solution) is not clear and other solutions from using different models might result in similarly good fits. The relative frequency of the mass of the most-massive star, \mmax, they derive for clusters of about 10$^3$\msun\,is shown as a dashed histogram in Fig.~\ref{fig:mmaxfreq}. When they use the \mMr\,as a truncation limit for clusters of about 10$^3$ $M_\odot$, a \mmax\,of 35 $M_\odot$ shouldn't be exceeded, but clearly it is. In this regard two statements by \citet{ACC13} are inconsistent. Firstly, \citet{ACC13} used SLUG code models of the SEDs of 10$^3$ \msun\,clusters and scaled these to fit the SEDs of the observed clusters which extend up to several 10$^4$. So when using the scaled results for clusters between 500 to 9000 \msun\,based on the 10$^3$ models the \mMr\,should have been used within the same limits. Also it is not clear why the SLUG (and the {\sc Starburst99} comparison) models where used with a Kroupa-IMF \citep{Kr02} but the truncated pseudo 'IGIMF' models were calculated with a single slope Salpeter IMF \citep{Sal55}. Secondly, the study considers clusters with ages up to 8 Myr. How any cluster with an age above 5 Myr can have any stars above $\approx$60 $M_\odot$ is unclear as these stars should have demised before that age. And while clusters of such ages might have had high-mass stars no trace of them can be present in the photometric data of the clusters.

Furthermore, the SLUG code itself does not actually use the stochastic sampling the authors claim. As described in \citet{DFK12}, clusters are chosen by mass from a mass function and then they are randomly filled with stars to achieve this cluster mass. This method is called 'mass-constrained sampling' in \citet{WK05b} because it introduces a bias to the galaxy-wide IMF which results from adding up such clusters to this IGIMF \citep[see eq.\,4.66 in][]{KWP13}, while real random sampling would preserve the IMF. This is visualised in Fig.~\ref{fig:constrained}. There the solid line shows the input canonical IMF (which is identical to the IGIMF when using random sampling), the dashed (red) line is the resulting IGIMF when using mass-constrained sampling to construct 2.5 $\times$ 10$^7$ star clusters randomly taken from an embedded cluster mass function (ECMF), $\xi(M_\mathrm{ecl}) \propto M_\mathrm{ecl}^{-\beta}$, between 5 and 10$^6$\msun\,with $\beta$ = 2.35. The dotted (blue) line depicts the resulting IGIMF using sorted sampling for the same number of star clusters and with the same ECMF as for mass-constrained sampling and the dash-dotted line is the IGIMF using the analytical \mMr. While the IGIMF derived from sorted sampling is as good as identical to the analytical IGIMF, neither mass-constrained sampling nor random sampling reproduce neither the analytical IGIMF nor the canonical IMF.

To show that the assumption of using the \mMr\,as a truncation limit in order to model realistic stellar populations is wrong, the sample of Milky Way and Magellanic Cloud star clusters by \citet{WKP13} is used. Applying the same limits as \citet{ACC13} to sample clusters with a mass of 10$^3$\msun, all clusters between 500 to 4000\msun\,are taken from table A1 in \citet{WKP13} and the relative frequency of their \mmax\,values is calculated. These data are shown as a solid histogram in Fig.~\ref{fig:mmaxfreq}. The error bars for these data are Poisson uncertainties by taking into account the lower and upper mass limits of the \mmax\,values of the clusters.

The $\chi^2$ value between the two data sets (red dashed vs solid black) is 0.971 (reduced $\chi_\mathrm{red}^2$ =  0.0971). For the eleven bins (10 degrees of freedom) this means that there is more than a 95\% probability that any difference between the two samples is just pure chance. Employing a KS-test to evaluate the hypothesis that the \citet{ACC13} sample stems from the same distribution as the \citet{WKP13} sample arrives at a similar result. In order to reject this hypothesis at a level of $p(0.001)$, the distance between the curves, $D$, would need to be larger than 0.87 and to reject at a level of $p(0.1)$, $D$ needs to be larger than 0.55 but $D$ is 0.11. This puts the probability for this number of data points that both samples stem from the same distribution at the 99.9\% level. As the Milky Way sample has been shown by \citet{WKB09} and \citet{WKP13} to follow the \mMr, it is obvious that the \citet{ACC13} conclusion that the \mMr\,does not exist is incorrect.

A further test of the \citet{ACC13} results is performed by means of Monte-Carlo simulations. 10$^6$ star clusters are randomly taken from an ECMF, with a slope of $\beta$ = 2 and a lower limit of 5\msun\,and an upper limit of $10^5$\msun. The clusters between 500 and 4000\msun\,are assigned most-massive stars (\mmax) by using the fit to the numerical solution of eqs.~\ref{eq:mmr1} and \ref{eq:mmr2} \citep[eq.~10 in][]{PWK07}. This approach gives identical results as optimal sampling. The relative frequency of \mmax\,for the Monte-Carlo approach is shown as a dotted histogram in Fig.~\ref{fig:mmaxfreq}. Comparing this distribution with the 'observed' results for NGC 4214 gives a $\chi^2$ of 0.464 (reduced $\chi_\mathrm{red}^2$ = 0.0464) - again well in the 95\% confidence regime of where both samples are likely drawn from the same distribution. Using the KS-test for the hypothesis that the \citet{ACC13} sample is from the same distribution as the Monte-Carlo sample gives a $D$-value of 0.29. As the $p$-values are the same as for the first KS-test, both samples are indistinguishable on a more than 99.9\% confidence level.

\subsection{Ionising luminosity}
\label{sub:ionphot}

\begin{figure}
\begin{center}
\includegraphics[width=8cm]{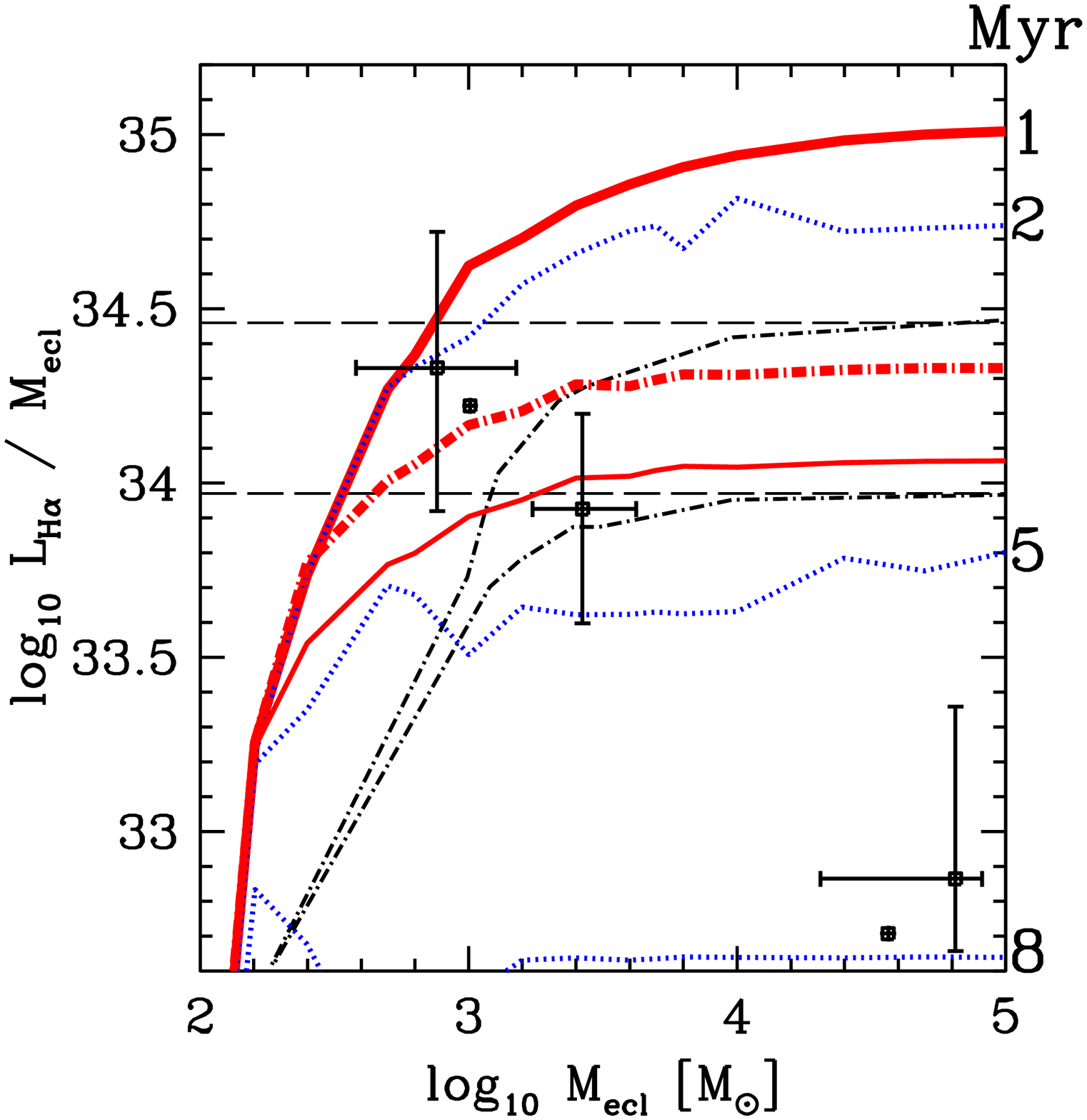}
\vspace*{-1.5cm}
\caption{H$\alpha$ luminosity weighted by cluster mass,  L$_{H\alpha}$/\mecl, versus cluster mass, \mecl. The boxes with and without error bars as well as the black dash-dotted lines and the long-dashed horizontal lines are from \citet{ACC13}. The upper black dash-dotted line is supposed to be for models with averaged ages between 2 and 5 Myr and assuming the \mMr\,is an upper truncation limit, while the lower dash-dotted line is for ages between 2 and 8 Myr and also assumes the \mMr\,is an upper limit on the stellar mass. The upper horizontal long-dashed line is the expected L$_{H\alpha}$/\mecl\,ratio for a universal IMF at an age between 2 to 5 Myr, and the lower horizontal long-dashed line is the same for an age between 2 to 8 Myr. The topmost red thick solid line is here derived for 1 Myr old clusters with the \mMr\,as the upper limit as obtained by using optimal sampling, the blue dotted lines are for 2 Myr, 5 Myr and 8 Myr old clusters (from top to bottom). The red dash-dotted line is the mean when using 2 to 5 Myr old clusters and the red thin solid line is the mean for ages between 2 and 8 Myr. The here presented values use the \citet{MM03} stellar evolution models and the \citet{SNC02} stellar atmosphere models to derive the ionising flux of massive stars.} \label{fig:andrews}
\end{center}
\end{figure}

As a further proof against the \mMr\,\citet{ACC13} invoke the H$\alpha$ luminosity of the clusters, normalised by the cluster mass, \mecl. The data from their figure 5 are show in our Fig.~\ref{fig:andrews}. Because the NGC 4214 clusters in the lowest mass bin (box with error bars at $\log_{10}$(\mecl/\msun) $\approx$ 3) are above their expected relation when assuming the \mMr\,is a truncation limit (thin dash-dotted lines), they argue the \mMr\,can not be right. 

\begin{figure}
\begin{center}
\includegraphics[width=8cm]{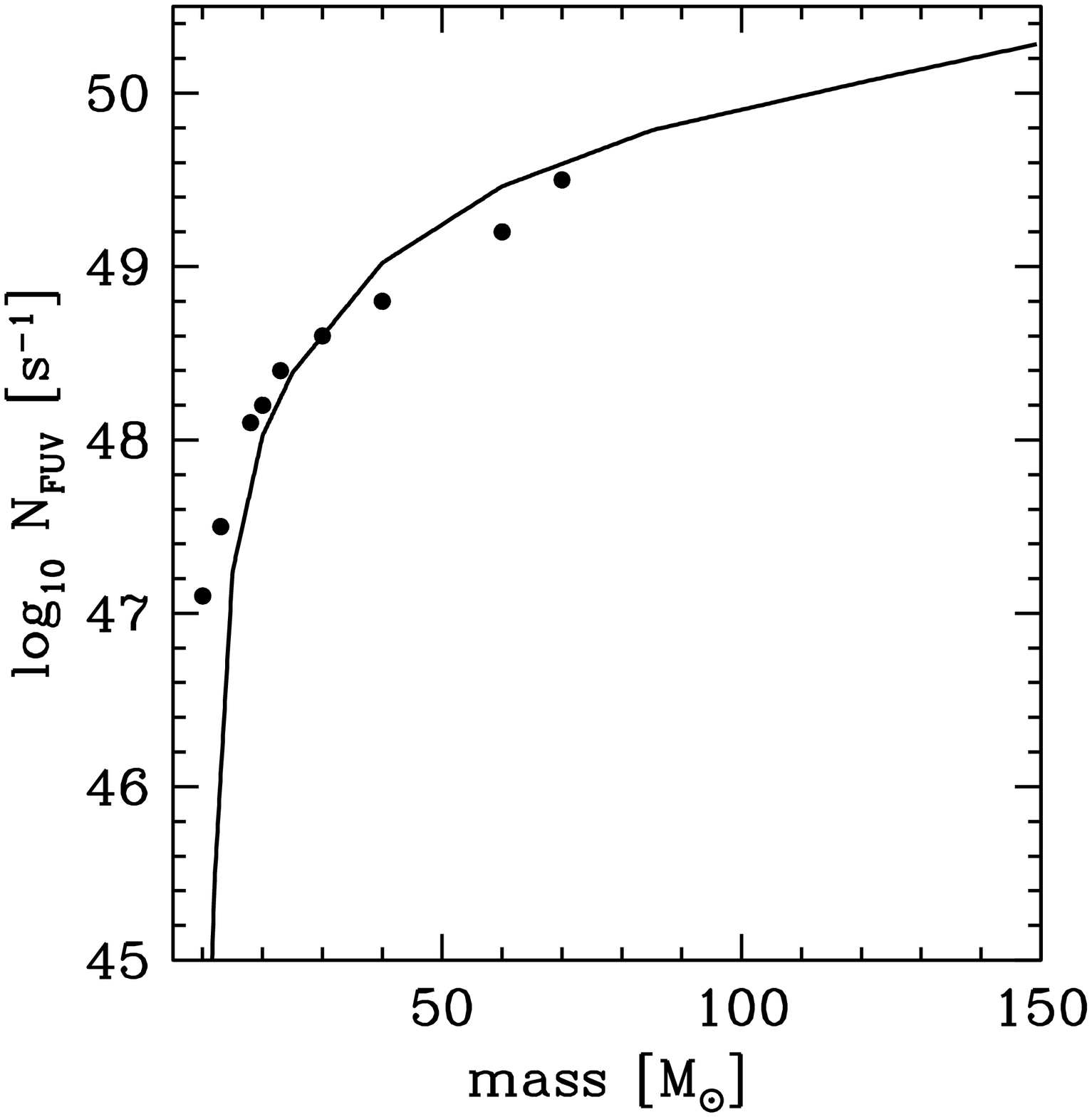}
\vspace*{-1.5cm}
\caption{Ionising photon flux in dependence of stellar mass. The black dots are from \citet{SP05} while the solid line is derived as described in the text.} \label{fig:ionphot}
\end{center}
\end{figure}

In order to investigate the impact of different sampling and the use of other stellar models on the L$_{H\alpha}$/\mecl\,values, 15 cluster masses between 100\msun\,and 10$^5$\msun\,in 0.2 dex logarithmic mass bins are populated with stars by optimal sampling \citep{KWP13}, which fulfils the analytical \mMr, with the software McLuster \citep{KMK11}. The ionising luminosity of the stars above 10\msun\,is calculated by using the solar metallicity rotating stellar models of \citet{MM03} and the stellar atmosphere models for O stars are from \citet{SNC02}. The flux below 912 \AA\,gives the ionising luminosity, $L_\mathrm{ion}$. This $L_\mathrm{ion}$ in units of erg s$^{-1}$ is used to derive the number of ionising photons, $N_\mathrm{ion}$,

\begin{equation}
N_{ion} = \log_{10}(L_\mathrm{ion}) + 10.5.
\end{equation}
Fig.~\ref{fig:ionphot} shows $N_\mathrm{ion}$ in dependence of stellar mass of individual stars. It can be seen that the here calculated values agree reasonably well with the numbers from table 15.1 from \citet{SP05}. This $N_\mathrm{ion}$ is then used to calculate $L_{H\alpha}$,

\begin{equation}
L_{H\alpha} = \mu \times N_\mathrm{ion} \times 3.0207 \cdot 10^{-12} \mathrm{erg s}^{-1} .
\end{equation}
The factor $\mu$ indicates the fraction of ionising photons which actually produce H$\alpha$ emission and is set to $\mu$ = 1. The resulting $L_{H\alpha}$ from each star in the model clusters is added up to calculate the total H$\alpha$ luminosity of the cluster and is divided by \mecl\,to get the L$_{H\alpha}$/\mecl\,values. This hash been done for fixed ages of 1, 2, 5 and 8 Myrs, as well as for 100 thousand ages between 2 and 5 Myr and 2 and 8 Myr to derive the averaged values. Here the L$_{H\alpha}$ were added first and then divided by the \mecl. The so derived values are plotted together with the the \citet{ACC13} data points and model lines in Fig.~\ref{fig:andrews}.

The upper and the middle blue dotted lines in the Figure mark clusters of 2 and 5 Myr, respectively and the lowest blue dotted line is for an age of 8 Myr, while the red thick dash-dotted line is the average between 2 and 5 Myr for the here used models and the thin red solid line is the average for 2 to 8 Myr. As can be seen in Fig.~\ref{fig:andrews}, the L$_{H\alpha}$/$M_\mathrm{ecl}$ ratios averaged over 2 to 5 Myr (upper black thin dash-dotted line) and 2 to 8 Myr (lower black thin dash-dotted line) fail to explain the observations in NGC 4214 mainly because the averaging hides the spread in L$_{H\alpha}$/$M_\mathrm{ecl}$ at such ages. Within the ranges given by the non-averaged values (our blue dotted lines in Fig.~\ref{fig:andrews}), all observations of NGC 4214 are readily explained also by clusters using the \mMr\,as a truncation limit.

The discrepancy in $L_{H\alpha}$ between the \mMr\,models by \citet{ACC13} and the ones presented here could be due to several factors. Different stellar evolution models used can certainly have a large impact. Contributing could be as well the lower limit chosen by \citet{ACC13} for stars to have H$\alpha$ emission and the use of a single slope Salpeter IMF for the truncated clusters in the \citet{ACC13} study as this reduces the number of stars above 8 \msun\,by about 50\%. The difference in metallicity used (in this work solar, in \citealt{ACC13} z = 0.004) could also be contributing to the discrepancy.

Furthermore, it has to be kept in mind that the models calculated here also use the \mMr\,as a truncation limit through the optimal-sampling procedure. Allowing for sorted sampling and using e.g.\,the one-$\sigma$ spread of the \mmax-\mecl\,observational data around the mean of sorted sampling would introduce an even larger spread in L$_{H\alpha}$ values. It is, however, clear that rejection of the \mMr\,based on the argument made by Andrews et al. is not correct..

\subsection{Photometric cluster masses}

Additionally, \citet{ACC13} employ a range of fairly small apertures for the photometry, which is used to derive the cluster masses. As stated in \citet{ACC13} a 3 pixel radius (1.74 pc at 3 Mpc) is used. But all clusters in NGC 4214 are between 2.2 and 7.5 Myr old. Gas expulsion from the embedded state will lead to considerable expansion and star loss of the clusters within a few Myr. \citet{KAH01}, \citet{BK08} and \citet{BK13} showed that already after 2 Myr the re-virialised core of a cluster typically has a radius between 1.5 and 2 pc. This has been observationally shown by \citet{BGG08} who found in a study of extragalactic star clusters that clusters with an age of 7 Myr have typical photometric core radii of about 1 pc. As such clusters are dynamically (or even primordially) mass segregated the photometric core radius of extragalactic star clusters is about a factor of 2  smaller than the real one because the massive stars dominate the light \citep{GG08}. Beyond this radius about 60-90\% of the initial stellar cluster mass is already lost \citep[see also][]{MK12}. This results in a serious underestimation of the initial mass of the cluster not only by the fact that most of the lost mass is outside the aperture but also as this mass is most-likely still close enough to be at least partially in the annulus outside the aperture used to correct for the background. Furthermore, this also means that the cluster mass is not age independent as claimed by \citet{ACC13} and \citet{CCL10}.

\section{Conclusions}
\label{se:conclusions}

\citet{ACC13} claim that the data obtained from the young starburst dwarf galaxy NGC 4214 falsifies the \mMr\,of \citet{WKB09} and with it the IGIMF of \citet{KW03} and \citet{KWP13}. As show in \S~\ref{se:ngc4214} this claim does not stand up to closer inspection. This is due to the following:

\begin{itemize}
\item The SFR of NGC 4214 is relatively high (SFR $\approx$ 0.2 \msun\,yr$^{-1}$) but the IGIMF effect (i.e.. the steepening of the galaxy-wide IMF with a deficit of massive stars in comparison to a canonical IMF which depends on the \mMr\,in star clusters) on ionising emissions and \mmax\,are only to be expected for SFRs below $\approx$ 0.05\msun\,yr$^{-1}$ (see figure 5 of \citealt{PWK07}). For galaxies with global 0.1 \msun\,yr$^{-1}$ $<$ SFR $<$ 5 \msun\,yr$^{-1}$ distributed over a fully populated embedded cluster mass function no difference between the relative population of massive stars to the one of the Milky Way is to be expected. 
\item The \mMr\,is used as a truncation limit for a randomly sampled IMF in \citet[][and as well in \citealt{FDK11}, \citealt{DFK12}]{ACC13}. This leads to inconsistencies with the observed \mmax\,values for Milky Way star clusters as can be seen in Fig.~\ref{fig:samplingmmax}. Sorted sampling, for example, avoids such inconsistencies.
\item Strikingly, the \mmax\,values from the best-fitting procedure of \citet{ACC13} show no trend of \mmax\,with \mecl. It is currently not possible to explain such behaviour by any known sampling procedure. However, when comparing the relative frequency of the \mmax\,values of the NGC 4214 sample for clusters with masses between 500 and 4000\msun\,with a sample of young Milky Way clusters in the same mass range, the samples are consistent with being from the same distribution at high confidence. In \citet{WKP13} it has been shown that the Milky Way sample shows a physical \mMr.
\item The H$\alpha$ luminosity to \mecl\,ratios given by \citet{ACC13} when using the \mMr\,as a truncation limit do not agree with values independently calculated here (Fig.~\ref{fig:andrews}) while our independent derivation reproduces literature values by \citet{SP05}. Using the here derived models explains all three cluster mass bins of \citet{ACC13} well even when using the \mMr\,as a truncation limit.
\end{itemize}

If the \mMr\,is applied as a truncation limit for a randomly sampled IMF in star clusters, the NGC 4214 data do not support the existence of a non-trivial \mMr. However, the analytical \mMr\,is an average value with observational data clustered above and below it. Using it as a truncation limit cuts off the portion of the distribution above the \mMr,\,therefore suppressing massive stars which would be expected even in the case that a physical \mMr\,does exist.

As can been seen above, the \citet{WKP13} sample of star clusters for the respective cluster mass range reproduces the \citet{ACC13} distribution of most-massive stars in NGC 4214 very well. It also follows that it is possible to reproduce the observed H$\alpha$ luminosities when applying the \mMr. The difference in the conclusions by \citet{ACC13} and this study are likely due to degeneracies between different sampling methods combined with the uncertainties of models for massive stars. In order to further constrain the \mMr\,a larger sample of well studied, preferably resolved, clusters is necessary, as well as in-depth studies of galaxies with very low SFRs \citep{WKP13}.

\section*{Acknowledgements}
We thank Daniela Calzetti for helpful discussions and suggestions. This work has been supported by the Programa Nacional de Astronom{\'i}a y Astrof{\'i}sica of the Spanish Ministry of Science and Innovation under grant AYA2010-21322-C03-02.
\begin{appendix}

\section{The canonical IMF}
\label{app:IMF}
The following two-component power-law stellar IMF is used throughout the paper:

{\small
\begin{equation}
\xi(m) = k \left\{\begin{array}{ll}
k^{'}\left(\frac{m}{m_{\rm H}} \right)^{-\alpha_{0}}&\hspace{-0.25cm},m_{\rm
  low} \le m < m_{\rm H},\\
\left(\frac{m}{m_{\rm H}} \right)^{-\alpha_{1}}&\hspace{-0.25cm},m_{\rm
  H} \le m < m_{0},\\
\left(\frac{m_{0}}{m_{\rm H}} \right)^{-\alpha_{1}}
  \left(\frac{m}{m_{0}} \right)^{-\alpha_{2}}&\hspace{-0.25cm},m_{0}
  \le m < m_\mathrm{max},\\ 
\end{array} \right. 
\label{eq:4pow}
\end{equation}
\noindent with exponents
\begin{equation}
          \begin{array}{l@{\quad\quad,\quad}l}
\alpha_0 = +0.30&m_\mathrm{low} = 0.01 \le m/{M}_\odot < m_\mathrm{H} = 0.08,\\
\alpha_1 = +1.30&0.08 \le m/{M}_\odot < 0.50,\\
\alpha_2 = +2.35&0.50 \le m/{M}_\odot \le m_\mathrm{max}.\\
          \end{array}
\label{eq:imf}
\end{equation}}
\noindent where $dN = \xi(m)\,dm$ is the number of stars in the mass interval $m$ to $m + dm$. The exponents $\alpha_{\rm i}$ represent the standard or canonical IMF \citep{Kr01,Kr02,KWP13}. For a numerically practical formulation see \citet{PAK06}.

The advantages of such a multi-part power-law description are the easy integrability and, more importantly, that {\it different parts of the IMF can be changed readily without affecting other parts}. Note that this form is a two-part power-law in the stellar regime, and that brown dwarfs contribute about 1.5 per cent by mass only and that brown dwarfs are a separate population \citep[$k^{'} \approx \frac{1}{3}$,][]{TK07,TK08}.

The observed IMF is today understood to be an invariant Salpeter/Massey power-law slope \citep{Sal55,Mass03} above $0.5\,M_\odot$, being independent of the cluster density and metallicity for metallicities $Z \ge 0.002$ \citep{MH98,SND00,SND02,PaZa01,Mass98,Mass02,Mass03,WGH02,BMK03,PBK04,PAK06}.  Furthermore, un-resolved multiple stars in the young star clusters are not able to mask a significantly different slope for massive stars \citep{MA08,WK07c}. \citet{Kr02} has shown that there are no trends with present-day physical conditions and that the distribution of measured high-mass slopes, $\alpha_3$, is Gaussian about the Salpeter value thus allowing us to assume for now that the stellar IMF is invariant and universal in each cluster. There is evidence of a maximal mass for stars \citep[$m_{\rm max*}\,\approx\,150\,M_{\odot}$,][]{WK04}, a result later confirmed by several independent studies \citep{OC05,Fi05,Ko06}. However, according to \citet{CSH10} $m_\mathrm{max*}$ may also be as  high as 300 $M_\odot$ \citep[but see][]{BK12}. \citet{DKP12}, \citet{MKD10} uncovered a systematic trend towards top-heaviness (decreasing $\alpha_3$) with increasing star-formation rate density.

\end{appendix}

\bibliography{mybiblio}
\bsp
\label{lastpage}
\end{document}